\documentclass[
  aps,
  prab,
  12pt,
  onecolumn,
  a4paper,
  notitlepage,
  groupedaddress,
  nofootinbib
]{revtex4-2}

\usepackage{amsmath,amssymb,bm}
\usepackage{booktabs}
\usepackage{graphicx}
\usepackage{placeins}
\usepackage{microtype}
\usepackage{siunitx}
\usepackage[colorlinks=true,allcolors=black]{hyperref}

\newcommand{\dd}{\mathop{}\!\mathrm d}
\newcommand{\avg}[1]{\left\langle #1\right\rangle}

\AddToHook{env/figure/begin}{\linespread{1.04}\selectfont}
\AddToHook{env/table/begin}{\linespread{1.04}\selectfont}

\begin{document}
\raggedbottom
\clubpenalty=10000
\widowpenalty=10000

\title{A multiscale composition method for tracking electron bunches from periodically nanostructured photocathodes}

\author{Dmitry Bazyl}
\email[Corresponding author: ]{dmitry.bazyl@desy.de}
\author{Igor Zagorodnov}
\affiliation{Deutsches Elektronen-Synchrotron DESY, Notkestrasse 85, 22607 Hamburg, Germany}

\date{September 10, 2026}

\begin{abstract}
Nanoholes and other periodic structures on a photocathode can enhance
emission by increasing optical absorption and concentrating light. The
curvature of a nanohole also alters electron emission directions and
the local accelerating field. For the injector considered here,
resolving holes of 200~nm nominal width across a laser spot 0.87~mm
in diameter is computationally not feasible. We propose a multiscale
composition method for tracking electron bunches from nanostructured
photocathodes through rf injectors using conventional beam dynamics codes.
For a cathode patterned with a $41\times41$ array of nanoholes, the
method reproduces the transverse phase space obtained by resolving
every hole in WarpX. The projected emittances agree to within 0.1\% in both
planes. We then apply this method to calculate the dynamics of a 100~pC
bunch in an rf photoinjector with solenoid focusing.
For a flat cathode with smooth emission across
the same laser spot, the central-slice emittances at 5~m are about
15\% above the nominal thermal emittance. The predicted central-slice
emittances for the nanostructured cathode exceed those for the flat
cathode by 13\% horizontally and 19\% vertically. The vertical emittance
in this slice is modestly larger than the horizontal. No periodic modulation
is visible in the transverse phase-space projections of the whole bunch.
\end{abstract}

\maketitle

\section{Introduction}

Nanostructured photocathodes can enhance nonlinear photoemission by localizing the optical field~\cite{Polyakov2013,Pierce2023}. Their geometry
can also increase the transverse momentum spread through variations in the local emission direction and extraction
field~\cite{Pierce2023,Zhang2015,Gevorkyan2018}. Coupled calculations of optical absorption and electron transport have predicted quantum efficiency and intrinsic emittance for patterned semiconductor cathodes~\cite{Jiang2021,Popov2026}. In a photoinjector, the emitted distribution subsequently evolves under the applied fields and the space-charge field of the complete bunch. Predicting the effect of nanostructuring on the accelerated beam therefore requires coupling the dynamics near the cathode to the collective evolution of the full bunch.

Photoinjectors can operate with laser radii of several hundred micrometres
and, in some cases, rms spot sizes approaching a
millimetre~\cite{Graves2014,AngalKalinin2020,Xu2022}. For example,
applying the \SI{747}{nm}-pitch
nanohole array of Li \textit{et al.}~\cite{Li2013} over a circular
\SI{300}{\micro m}-diameter spot would illuminate approximately
$1.3\times10^5$ holes. To calculate the beam dynamics directly, we would
have to resolve the fields around every hole and the collective field of
the emitted bunch, then track the electrons through the injector.
This requires nanoscale resolution across the patterned area during
emission. Directly resolving the surface throughout such a calculation
is not computationally feasible at the injector scales considered here.

Previous work connected nanostructured emission to rf gun tracking using prescribed or replicated particle distributions. Nanotip beamlets were tracked through an rf gun and a phase-space-exchange beamline~\cite{Graves2012}.  Using WARP~\cite{Friedman2014}, Lueangaramwong \textit{et al.}~\cite{Lueangaramwong2017,LueangaramwongAAC2017} calculated
particle distributions from a single plasmonic nanohole, replicated them over small arrays, and tracked them with ASTRA~\cite{Floettmann2017} or IMPACT-T~\cite{Qiang2006,Qiang2009}. The authors examined whether the imposed beamlet patterns persisted during acceleration.

To extend local calculations to the full illuminated area, Bulgacheva \textit{et al.} proposed scaling them to a
macroscopic laser spot and extending the tracking through the electron gun~\cite{Bulgacheva2024TEMF}. Subsequent work for the European XFEL superconducting rf photoinjector coupled optical absorption and spatially resolved photoemission to particle tracking near the cathode~\cite{Bazyl2025Cathodes,Bulgacheva2026}.

A practical starting point is to reduce the computational domain to one period of the square
nanohole array. With transverse periodic boundary conditions, the hole
and its emission repeat without end in both directions along the cathode.
In this periodic calculation, the electrons experience the space-charge
field of an infinite array. Copying their distribution across a finite laser spot
preserves the effect of the nanohole on their momenta but does not
account for the space-charge impulse they would acquire near the cathode
in a bunch emitted across that spot.

In order to calculate beam dynamics of a bunch emitted from a nanostructured photocathode in an rf injector,
we combine particle tracking at two scales. First, we project the emission pattern
of each hole onto a flat cathode, retaining the charge profile across the
laser spot. From this \emph{projected-source planar emitter}, the bunch
can be tracked with self-consistent space charge in WarpX~\cite{Vay2018}
or IMPACT-T~\cite{Qiang2006,Qiang2009} without resolving individual holes.
To estimate how the surface geometry changes electron motion on the scale of one hole, we compare
two periodic WarpX calculations with the same pitch, emission samples
and charge per period. In one, the emitting surface is flat while in the
other, it contains a nanohole. At a common plane above the cathode,
we determine how the hole changes each electron's position, momentum
and arrival time. We repeat the comparison at several charges per period.
For each period in the laser spot, we evaluate these changes at the
charge emitted by that period and add them to the corresponding particle
records from the projected-source planar emitter. During a long pulse,
early electrons can leave through the upper boundary of the periodic
domain while emission continues.
To include the retarding force on electrons emitted later, we reconstruct
the field at the cathode from the electrons that have left and their image charge.

Section~\ref{sec:composition} describes how to construct the bunch for
a conventional rf injector code from a prescribed photoemission distribution
on a periodically structured cathode. In Sec.~\ref{sec:model}, we apply the
composition method to a $41\times41$ array of Gaussian nanoholes and recover
the projected emittances obtained by resolving every hole in WarpX.
We also examine the effects of particle count and mesh resolution in
IMPACT-T. In Sec.~\ref{sec:injector_application}, we
consider a 100~pC source spanning more than a million holes and
track the composed bunch through an L-band rf gun and solenoid. We
compare its slice emittances with those for the projected-source planar
emitter and for a flat cathode with smooth emission across the same spot.

\section{Composition method for periodically structured photocathodes}
\label{sec:composition}
\label{sec:state_composition}

\begin{figure}[!htbp]
 \centering
 \includegraphics[width=0.93\linewidth,trim=0bp 24bp 0bp 0bp,clip]{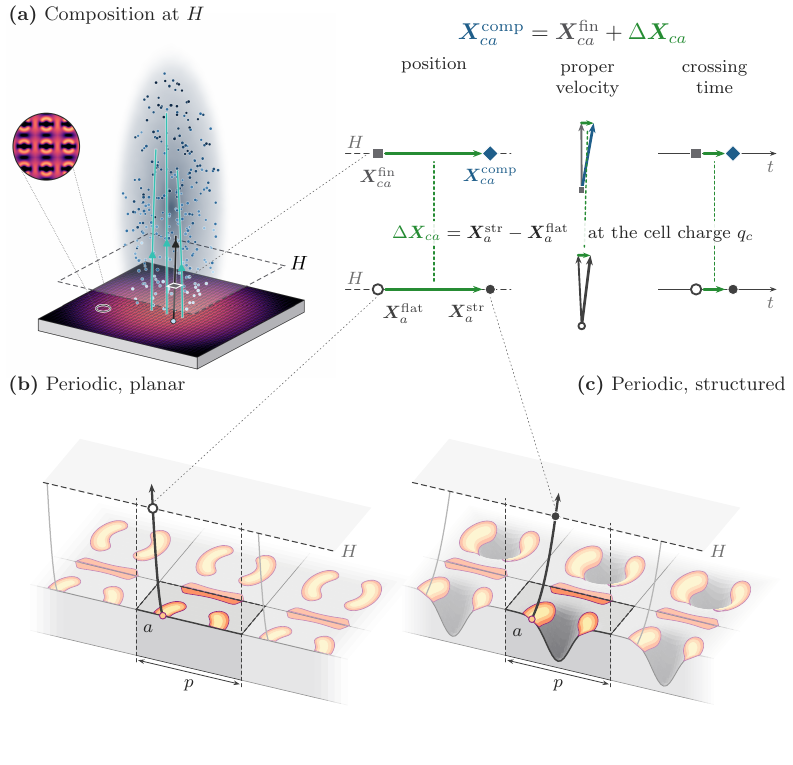}
 \caption{Composition method for rf-injector tracking of a bunch from
 a structured photocathode.
 (a) For sample $a$ in cell $c$ of the projected-source planar emitter,
 the square marks the particle's crossing state at $H$. Adding the periodic
 difference at the cell charge $q_c$ (green) gives the composed crossing
 state (diamond). Teal arrows show the electric field of previously
 emitted electrons and their image charge in the cathode.
 The yellow points labelled $a$ mark the same emission sample at birth
 on (b) the planar periodic emitter and (c) the structured periodic emitter.
 The open circle in (b) and the filled circle in (c) mark the crossing
 states at $H$.}
 \label{fig:composition}
\end{figure}

For injector tracking, we construct the electron distribution at a plane
above the nanohole array. On a flat cathode we can retain the hole emission
pattern across the laser spot and track the emitted bunch with self-consistent
space charge [Fig.~\ref{fig:composition}(a)]. To each particle's transverse
position, momentum and arrival time we add the change from a flat surface
to a nanohole in one repeating period [Fig.~\ref{fig:composition}(b,c)]. A \emph{cell} is one square of the
cathode lattice, of side $p$ equal to the spacing between adjacent hole
centers.

Within each cell, emission is concentrated on the bright arcs
in the enlargement. We sample the transverse birth positions from this
local photoemission density. Each tracked particle is a macroparticle
of charge $-e w_i$, where $e>0$ is the elementary charge and the weight
$w_i>0$ is the equivalent number of electrons it represents. The weights
reproduce the variation in emitted charge across the laser spot.

\begin{figure}[!htbp]
 \centering
 \includegraphics[width=\linewidth]{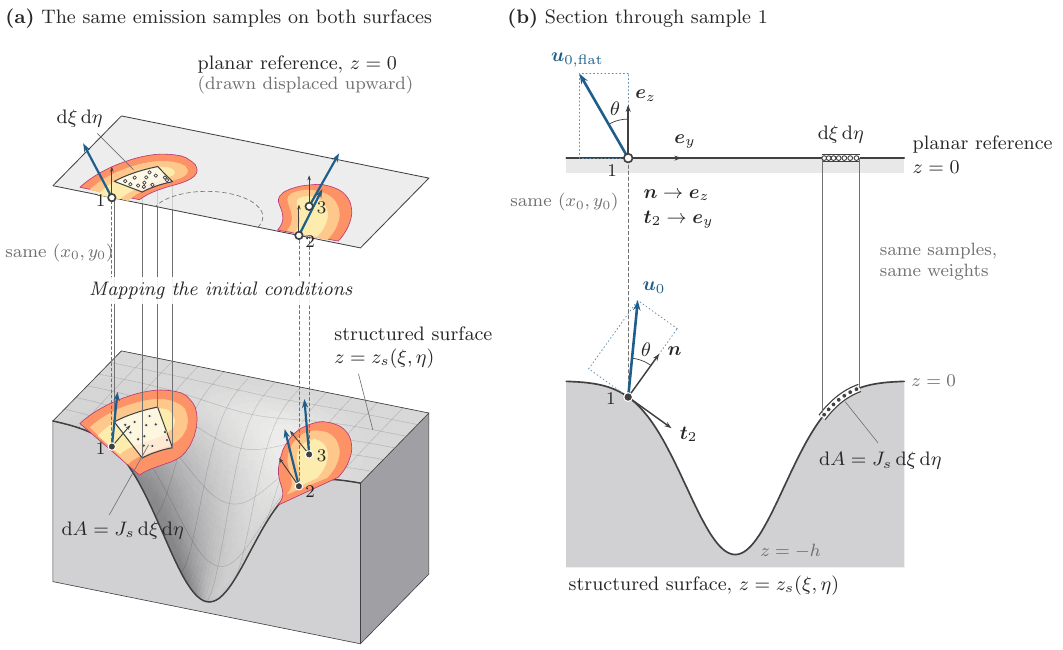}
 \caption{Projection of emission samples from a nanohole onto a plane.
 (a) Matching numbers identify the same emission sample $a$ on the two
 surfaces. The plane is drawn above its physical position. Sampling accounts
 for the emitting area, so projection leaves the weights unchanged.
 (b) For the pair marked 1, the launch proper velocity has the same
 components in the local surface frame on the hole and the Cartesian frame on the
 plane [Eq.~\eqref{eq:flat_mapping}]. The tangent $\bm t_1=\bm e_x$
 is perpendicular to the section.}
 \label{fig:source_mapping}
\end{figure}

For each \emph{emission sample} $a$ on the structured cathode, we place
its planar counterpart at the same transverse position $(x_0,y_0)$ but
at height $z=0$ [Fig.~\ref{fig:source_mapping}(a)]. The birth time,
kinetic energy and weight are unchanged. For the launch direction we use
the same polar and azimuthal angles in the local frame of each surface
[Fig.~\ref{fig:source_mapping}(b)]. An electron emitted along an inclined
normal on a nanohole therefore leaves along $z$ on the flat cathode.
The flat cathode and the emission pattern projected onto it across the
laser spot form the \emph{projected-source planar emitter}. We track the
emitted bunch with self-consistent space charge, including the field of
the charge it induces on the flat cathode (the image-charge field).
With the projected-source planar emitter we omit both the effect of the
surface slope on the launch directions in the laboratory frame and the
distortion of the electric field by the nanohole.

To isolate the effect of the nanohole geometry, we compare two infinite
arrays with the same pitch and emitted charge $q_j$ per cell.
In the \emph{planar periodic emitter}, the projected
emission pattern repeats across a flat cathode [Fig.~\ref{fig:composition}(b)].
In the \emph{structured periodic emitter}, each cell contains one nanohole
[Fig.~\ref{fig:composition}(c)]. For each emission sample $a$ we track
one particle from each of these two emitters.
We simulate one cell of each array. A particle that leaves through one
transverse side re-enters through the opposite side. With the cathode grounded,
we calculate the space-charge field of the electrons in the cell and
their copies throughout the array. To set the driving amplitude for both
cells at each time, we use the normal extraction field at a flat cathode
in the absence of emitted charge.

We compare the trajectories at a height $H$ above the surrounding flat
cathode where the geometrical modulation of the extraction field is small.
In the absence of emitted charge, its longest spatial harmonic has
wavelength $p$ and decays as $\exp(-2\pi z/p)$. In all three calculations
we record particle $i$ when it first crosses this plane moving away from
the cathode. The paired particles can arrive at different times, so we
include the arrival time along with the transverse position and proper
velocity in the crossing state
\begin{equation}
 \bm X_i(H)=(x_i,y_i,u_{x,i},u_{y,i},u_{z,i},t_i),
 \qquad
 \bm u_i=\frac{\bm p_i}{m_e}=\gamma_i\bm v_i.
 \label{eq:crossing_state}
\end{equation}
Here $\bm p_i$ and $\bm v_i$ are the mechanical momentum and velocity,
$m_e$ is the electron rest mass and $\gamma_i$ is the Lorentz factor.
The arrival times are measured on the same clock, and the transverse
positions include any displacement across periodic boundaries. If both
particles from sample $a$ reach $H$, we subtract the crossing state for
the planar periodic emitter (open circle in Fig.~\ref{fig:composition}(b))
from the crossing state for the structured periodic emitter (filled circle
in (c)):
\begin{equation}
 \Delta\bm X_a(H;q_j)=
 \bm X_a^{\mathrm{str}}(H;q_j)-\bm X_a^{\mathrm{flat}}(H;q_j).
 \label{eq:periodic_difference}
\end{equation}
This difference includes the change in launch direction and the change in
motion under the extraction and space-charge fields.

For the same emission sample $a$, the square in Fig.~\ref{fig:composition}(a)
marks the crossing state $\bm X^{\mathrm{fin}}_{ca}(H)$ of the particle
emitted from cell $c$ of the projected-source planar emitter. We evaluate
the periodic difference for $a$ at the charge $q_c$ emitted by that cell, interpolating
between calculated charges where needed (Sec.~\ref{sec:cell_charge}).
Adding this correction (green arrows) to the crossing state for the
projected-source planar emitter gives the composed crossing state (diamond):
\begin{equation}
 \bm X^{\mathrm{comp}}_{ca}(H)=
 \bm X^{\mathrm{fin}}_{ca}(H)+\Delta\bm X_{ca}(H).
 \label{eq:composed_state}
\end{equation}
Applying the periodic difference to a cell within the laser spot assumes that the
macroscopic charge density and field vary little over one period and that
the local correction depends only weakly on the difference between the
macroscopic fields of the infinite array and the bunch emitted across the
laser spot. The corrected crossing states form the \emph{composed bunch}.
Each macroparticle represents the same number of electrons $w_i$ as its
counterpart in the projected-source planar emitter.

\subsection{Cell charge and particle sampling}
\label{sec:cell_charge}

The hole geometry repeats across the laser spot, but the charge emitted
per period varies, changing the space-charge forces near each hole.
Here $q_c$ and $q_j$ refer to the total charge emitted by a cell over
the pulse, expressed as a positive magnitude. Let $\Sigma_q(\bm R)$
denote the smooth envelope of the emitted charge magnitude per unit
area of the cathode plane. For cell $c$ with projected area
$\mathcal C_c$ and center $\bm R_c$, the envelope gives
\begin{equation}
 q_c\equiv\int_{\mathcal C_c}\Sigma_q(\bm R)\,\dd^2R
 \simeq p^2\Sigma_q(\bm R_c).
 \label{eq:slow_coordinate}
\end{equation}
We use this estimate of $q_c$ to evaluate the difference between the two
periodic crossing states [Eq.~\eqref{eq:periodic_difference}].
The particle weights of the projected-source planar
emitter can still follow the envelope within a cell
(Appendix~\ref{app:source}). When we retain fewer particles per cell,
we keep their emission-sample indices and choose weights that preserve
each cell's charge (Appendix~\ref{app:reduction}).

We repeat the calculations for the planar periodic emitter and the structured
periodic emitter at $N_q$ cell charges $q_j$ spanning the range across
the spot. For both emitters we divide $q_j$ equally among the same emission
samples. Varying the sample weights changes the charge per period
without changing the sampled birth conditions.
For a particle recorded at $H$ from the projected-source planar emitter,
we interpolate the periodic difference for its emission sample $a$
when it is available at both charges bracketing $q_c$:
\begin{equation}
 \Delta\bm X_{ca}(H)=
 \frac{q_{j+1}-q_c}{q_{j+1}-q_j}\Delta\bm X_a(H;q_j)
 +\frac{q_c-q_j}{q_{j+1}-q_j}\Delta\bm X_a(H;q_{j+1}).
 \label{eq:interpolated_difference}
\end{equation}
If the difference exists only at the nearer charge, we use that value
without interpolation. If it is absent at the nearer charge, we omit
the particle from the composed bunch (Appendix~\ref{app:response_assignment}).

We begin tracking each particle of the composed bunch at $H$ on the
first time step at or after its corrected crossing time. From that step
onward, its charge contributes to the downstream space-charge field
[Eq.~\eqref{eq:injector_activation}].

\subsection{Cathode field during long emission pulses}
\label{sec:long_emission}

\begin{figure}[!htbp]
 \centering
 \includegraphics[width=\linewidth]{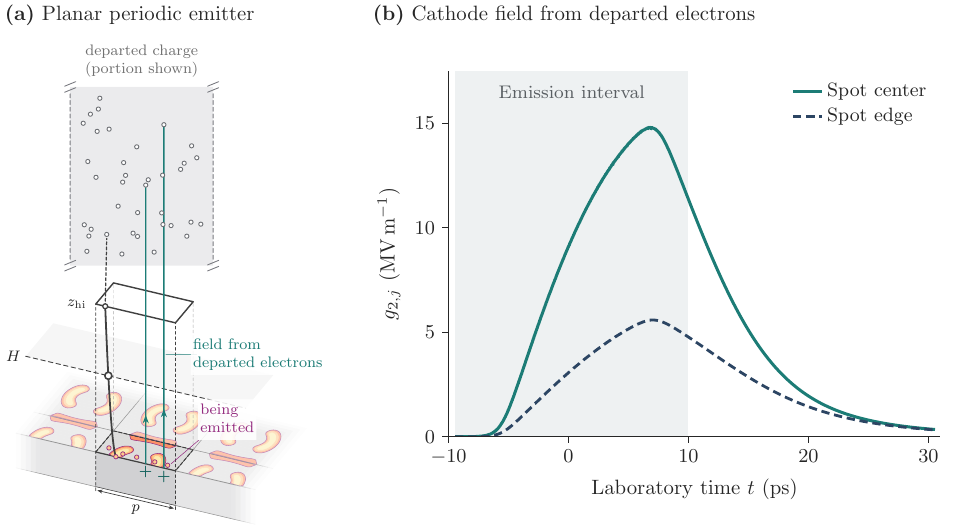}
 \caption{Cathode field of departed electrons during a long emission pulse.
 (a) For an electron from the planar periodic emitter, we record the
 crossing state at $H$ and include its charge in the field reconstruction
 only after it leaves through $z_{\mathrm{hi}}$.
 The electric field of the departed electrons and their image charge
 (teal arrows) opposes the applied rf field during emission.
 (b) Second-iteration field $g_{2,j}(t)$ at the center and outermost
 sampled radius of the injector spot (Sec.~\ref{sec:injector_application}).
 Shading marks the interval between the first and last sampled emission times.}
 \label{fig:departed_field}
\end{figure}

During a long pulse, the early electrons and their image charge exert a
retarding force on electrons still leaving the cathode. We include this
force directly when tracking the bunch from the projected-source planar
emitter [Fig.~\ref{fig:composition}(a)]. In a periodic cell, electrons
can leave through the upper boundary $z_{\mathrm{hi}}>H$ before emission
ends [Fig.~\ref{fig:departed_field}(a)]. They then cease to contribute to
the periodic space-charge field.

For the projected-source planar emitter, charge on both sides of
$z_{\mathrm{hi}}$ contributes to the cathode field. Each periodic
calculation already includes the field from charge still inside its
cell. To reconstruct the missing field, we record electrons from the
planar periodic emitter as they leave through $z_{\mathrm{hi}}$ and
follow their longitudinal motion beyond the boundary in the applied
rf electric field alone.
For each tabulated cell charge, these trajectories represent
electrons emitted over a range of radii in the laser spot. At their
calculated heights we distribute the corresponding charge over those
radii and include its planar image. Summing over the spot gives the
normal cathode field $g_j(t)$ at the radius where a cell emits charge
$q_j$ (Appendix~\ref{app:injector_field}). This field points into the
vacuum and opposes the applied rf field $-\mathcal E_{\mathrm{RF}}(t)$
during emission. The imposed normal field at the planar cathode is then
\begin{equation}
 E_{z,\mathrm{imp}}(q_j,t)=-\mathcal E_{\mathrm{RF}}(t)+g_j(t).
 \label{eq:long_pulse_field}
\end{equation}

The departure times and longitudinal momenta at $z_{\mathrm{hi}}$ depend
on the retarding field. Starting with $g_{0,j}(t)=0$, we repeat the calculation
for the planar periodic emitter using the field reconstructed from the preceding
iteration until it changes little at the sampled emission times.
For the rf injector in Sec.~\ref{sec:injector_application}, the
emission-weighted relative rms change fell from
\SI{1.5}{\percent} on the second update to \SI{0.048}{\percent} on the
third [Eq.~\eqref{eq:returned_field_difference}]. We therefore used
$g_{2,j}(t)$ [Fig.~\ref{fig:departed_field}(b)] in
Eq.~\eqref{eq:long_pulse_field} to scale the charge-free extraction field
of each periodic emitter. By using the same $g_{2,j}(t)$ for the
structured periodic emitter, we neglect the effect of the nanohole
geometry on the reconstructed field.

\section{Validation for a finite array of Gaussian holes}
\label{sec:model}
\label{sec:validation}

For an array of $41\times41$ nanoholes, we can resolve the electric
field around each hole and the space-charge field across the laser
spot with the same mesh spacing throughout the WarpX domain. At $H$ we
compare the \emph{resolved bunch}, tracked directly from this array, with
the composed bunch obtained for the same source by the method of
Sec.~\ref{sec:composition}.

\begin{figure}[!htbp]
 \centering
 \includegraphics[width=\linewidth]{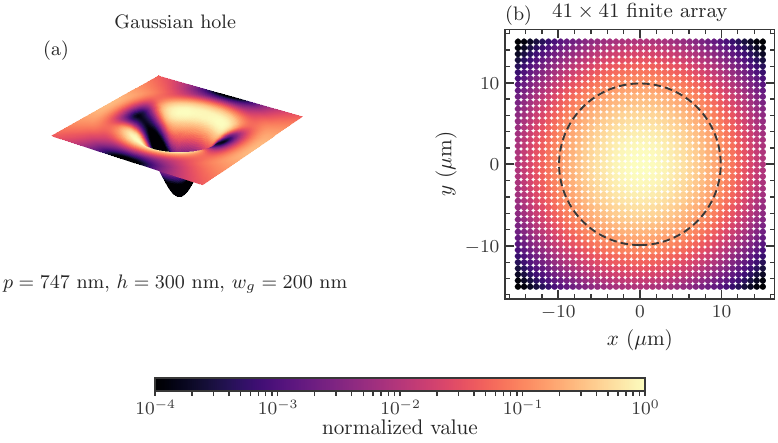}
 \caption{Validation source. (a) Projected emission density on one hole.
 (b) Emitted charge per cell across the array. The dashed circle marks
 the laser-intensity FWHM. The logarithmic scales are normalized separately.}
 \label{fig:method}
\end{figure}

We model the copper nanohole array studied by Li \textit{et al.}
with a lattice pitch of \SI{747}{nm}. Each hole has a Gaussian profile
of depth \SI{300}{nm} and nominal full width at half maximum (FWHM)
\SI{200}{nm}. Within one period
we approximate the optical intensity $I$ from the radial and angular
profiles in Fig.~4(b) and (c) of Ref.~\cite{Li2013}. For three-photon
emission the charge emitted per unit surface area is proportional to $I^3$,
so most electrons originate on two arcs on opposite sides of the hole
along $y$ [Fig.~\ref{fig:method}(a)].

The same arcs repeat across the array, but the charge per cell follows a
Gaussian envelope of rms width $\sigma_q=6.5p\simeq\SI{4.9}{\micro m}$
[Fig.~\ref{fig:method}(b)]. We choose a peak projected charge density of
\SI{50}{pC.mm^{-2}}, comparable to the pattern-averaged densities of
\SIrange{40}{50}{pC.mm^{-2}} in Fig.~3(b) of Ref.~\cite{Li2013}.
The total charge in our array is \SI{7.4}{fC}, or
\SI{7.9}{pC.mm^{-2}} averaged over the patterned square.

For the validation we use the \SI{150}{fs} laser-intensity FWHM reported
by Li \textit{et al.} and the \SI{35}{MV.m^{-1}} launch field from their
intrinsic-emittance calculation~\cite{Li2013}. We take the applied field
to be static and the laser pulse to be Gaussian. The cubic intensity
dependence then gives an emitted-current FWHM of \SI{87}{fs}
(\SI{37}{fs} rms). We sample birth times over an interval of about \SI{290}{fs}.
The initial kinetic energies are uniform from 0 to \SI{1}{eV}, and
the directions are isotropic over the outward hemisphere. We sample
8192 combinations of position within a cell, birth time, energy and
local emission angles with a scrambled Sobol sequence~\cite{Sobol1967}.
The same combinations are used in every cell and in both periodic
emitters. The particle weights follow the charge envelope across the
spot and are identical in the resolved array and the projected-source
planar emitter (Appendix~\ref{app:source}).

For the resolved array and the projected-source planar emitter we
use identical $2880\times2880\times192$ meshes. At the same spacing
of $p/64\simeq\SI{12}{nm}$, we calculate the periodic difference on
a $64\times64\times160$ mesh at eight cell charges. The upper
boundaries lie above the \SI{800}{nm} observation plane, at
\SI{1.49}{\micro m} in the periodic cells and \SI{1.87}{\micro m}
in both $41\times41$ calculations. With the cathode grounded, we fix
the upper potential to give the prescribed extraction field in the
charge-free planar geometry.
Tracking electrons from the resolved array for \SI{1.4}{ps} with a
Poisson solve every \SI{1}{fs} took 87 node hours (Appendix~\ref{app:numerics}).

\begin{figure}[!htbp]
 \centering
 \includegraphics[width=0.96\linewidth]{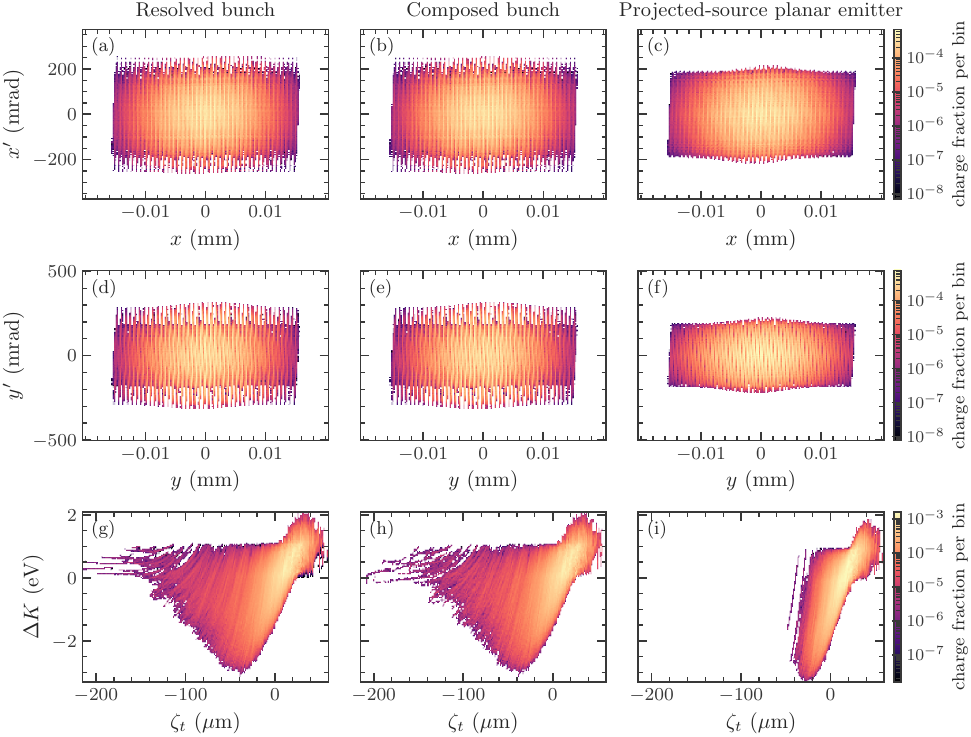}
 \caption{WarpX phase space at $H=\SI{800}{nm}$ for particles
 transmitted in all three calculations, with their weights from the resolved bunch.
 The slopes are $x'=u_x/u_z$ and $y'=u_y/u_z$.
 In (g--i), $\zeta_t=-c(t-\bar t_{\rm ref})$ and
 $\Delta K=K-\bar K_{\rm ref}$ are relative to the mean arrival time
 and kinetic energy of the resolved bunch.
 Negative $\zeta_t$ denotes later arrival. Color gives the charge fraction per bin.}
 \label{fig:phase_space}
\end{figure}

At $H=\SI{800}{nm}$ the angular spread increases most in $y$, along
the tilt of the surface normal at the emitting arcs
[Figs.~\ref{fig:method}(a) and~\ref{fig:phase_space}(a--f)].
The vertical projected normalized emittance of the resolved bunch is
\SI{5.9}{nm.rad}, \SI{42}{\percent} above the value for the projected-source
planar emitter. The increase in $x$ is \SI{9.7}{\percent}.
The projected emittances of the composed and resolved bunches agree to
within \SI{0.05}{\percent} in both planes.

For the same set of emission times, the rms arrival-time spread at
$H$ increases from \SI{51}{fs} for the
projected-source planar emitter to \SI{88.6}{fs} for the resolved
bunch. For the composed bunch the rms spread is \SI{88.4}{fs}, and the
tail of late arrivals agrees with that of the resolved bunch
[Fig.~\ref{fig:phase_space}(g--i)].
In the resolved array, about \SI{0.25}{\percent} of the emitted charge
returns to the cathode. We find the same fraction by composition when
we sum the charge of particles whose counterparts from the structured
periodic emitter return to the cathode at the nearest tabulated cell
charge (Appendix~\ref{app:response_assignment}).

\begin{figure}[!htbp]
 \centering
 \includegraphics[width=0.95\linewidth]{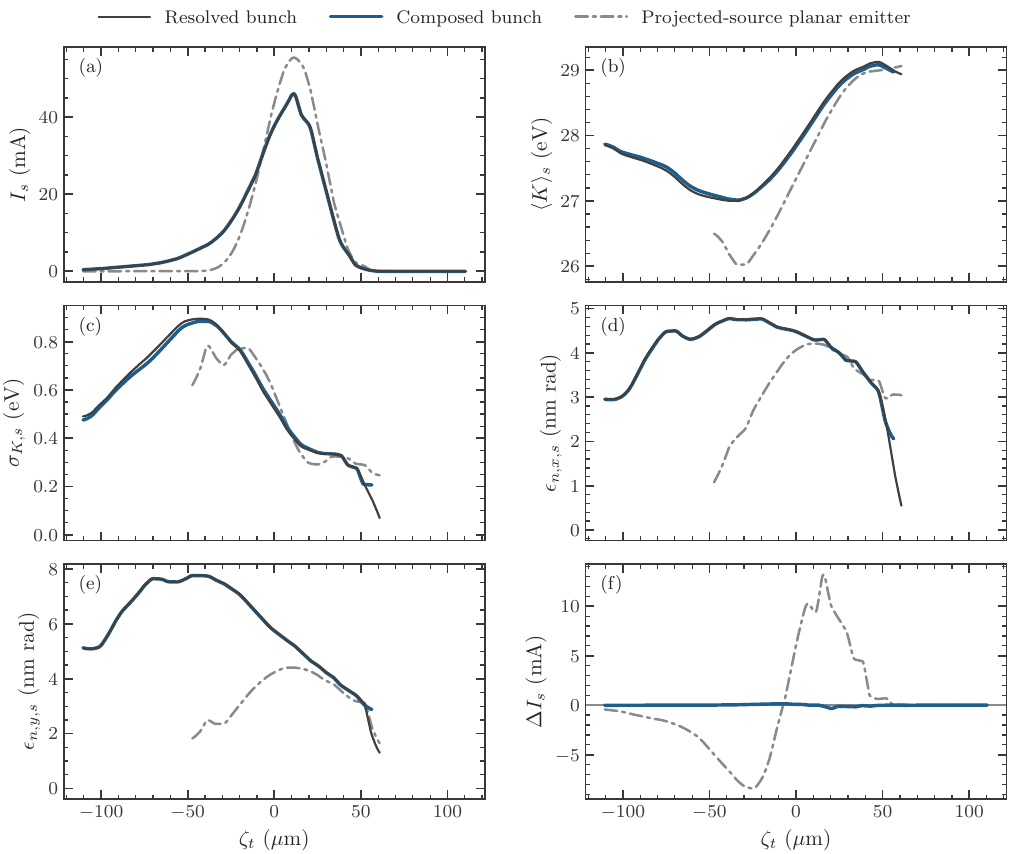}
 \caption{WarpX slice profiles at \SI{800}{nm} for the particles and
 arrival coordinate $\zeta_t$ of Fig.~\ref{fig:phase_space}. Panel (f) shows the current difference
 relative to the resolved bunch.}
 \label{fig:slices}
\end{figure}

The resolved and composed current profiles nearly coincide from the
peak into the tail [Fig.~\ref{fig:slices}(a)]. Both peak at
\SI{46}{mA}. For the projected-source planar emitter the same
matched charge of \SI{7.36}{fC} arrives over a shorter interval
and the peak reaches \SI{55}{mA}. The resolved and composed emittance
profiles agree in both planes, including the late slices where the
vertical emittance exceeds the horizontal [panels (d,e)]. For the
slice rms energy spread, the difference from the resolved profile over
equal-charge slices falls from \SI{31}{\percent} for the projected-source
planar emitter to \SI{2.0}{\percent} after composition.

The narrow charge envelope of the $41\times41$ array makes this a
demanding test of the composition method. Along a lattice axis the
density falls by \SI{14}{\percent} in one period, from $6p$ to $7p$
from the spot center, whereas neighboring cells in the periodic
calculation emit equal charges. Using the correction at each cell's
own charge nevertheless reproduces the transverse broadening and the
tail of late arrivals at \SI{800}{nm}. At this height and in an
independent composition at \SI{400}{nm}, the projected emittances and
the slice-emittance profiles agree with direct WarpX tracking to within
\SI{0.1}{\percent} in both transverse planes (Table~\ref{tab:planes}).

\begin{table}[!htbp]
 \caption{Relative differences between the composed and resolved WarpX
 bunches. Whole-bunch moments include all transmitted particles. For
 the slice profiles we compare the same equal-charge groups, ordered
 by arrival time in the resolved bunch, using Eq.~\eqref{eq:slice_error}. The
 emittance bounds apply separately in $x$ and $y$.}
 \label{tab:planes}
 \centering
 \small
 \renewcommand{\arraystretch}{1.12}
 \begin{tabular*}{0.84\linewidth}{@{\extracolsep{\fill}}lcc@{}}
  \toprule
  & \multicolumn{2}{c}{Observation height $H$} \\
  \cmidrule(l){2-3}
  Quantity & \SI{400}{nm} & \SI{800}{nm} \\
  \midrule
  Rms bunch duration & $<1\%$ & $<1\%$ \\
  Rms energy spread & $<1\%$ & $1.3\%$ \\
  Projected normalized emittance & $<0.1\%$ & $<0.1\%$ \\
  Slice emittance profiles & $<0.1\%$ & $<0.1\%$ \\
  Slice energy-spread profile & $1.1\%$ & $2.0\%$ \\
  \bottomrule
 \end{tabular*}
\end{table}

\subsection{Space charge and particle sampling in IMPACT-T}
\label{sec:results}

\begin{figure}[!htbp]
 \centering
 \includegraphics[width=0.92\linewidth]{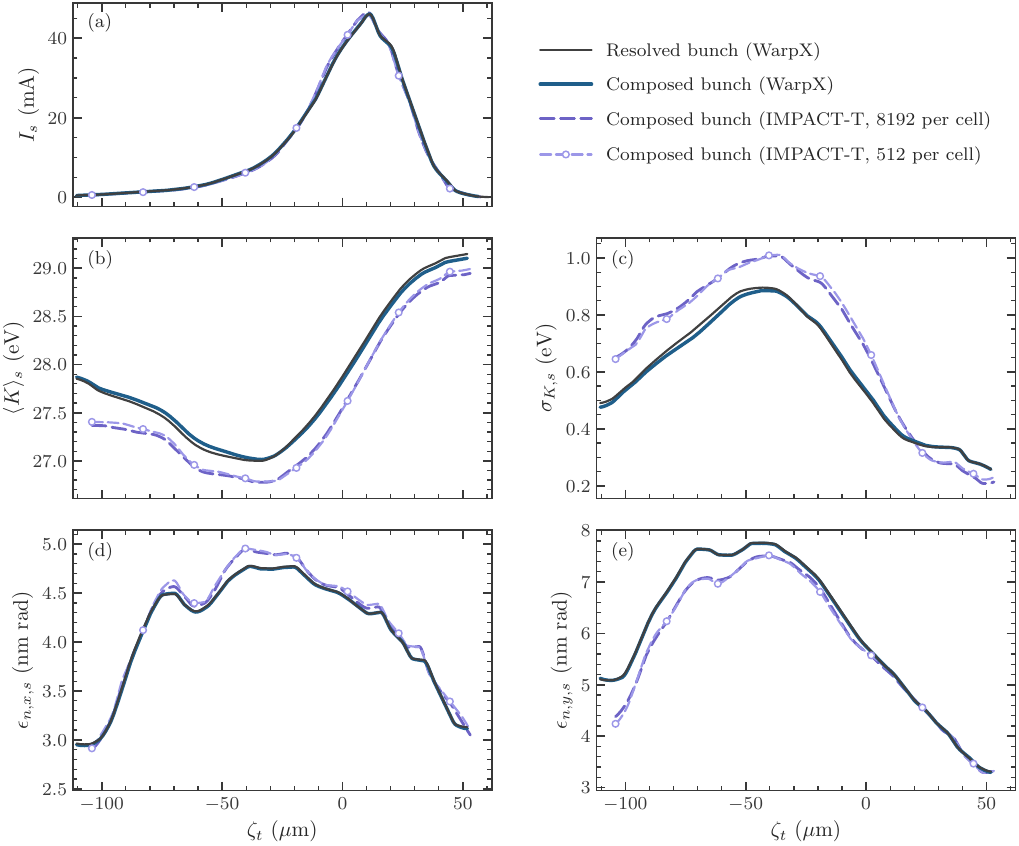}
 \caption{Slice profiles at \SI{800}{nm} for the resolved WarpX bunch
 and three composed bunches. Arrival times are relative to the mean
 arrival time of the resolved bunch. The periodic correction from WarpX is
 the same in all three compositions. The two IMPACT-T calculations differ
 in particle count and space-charge mesh.}
 \label{fig:practical}
\end{figure}

In Sec.~\ref{sec:injector_application} we track the bunch through an rf
injector with IMPACT-T~\cite{Qiang2006,Qiang2009,ImpactTCode}. Its
standard cathode model does not support emission from
a curved surface, so we track electrons from the projected-source planar
emitter in IMPACT-T and calculate the periodic correction in WarpX. For
the $41\times41$ array we can compare the slice profiles of the composed
bunch with those of the resolved bunch.
We do so first with all emission samples and then with fewer macroparticles
per cell and a coarser mesh in IMPACT-T (Fig.~\ref{fig:practical}).

Before applying the periodic correction, we track identical particles
from the projected-source planar emitter in both codes.
In IMPACT-T we calculate the field of the bunch and
its planar cathode image with open boundaries. In WarpX 26.06 the
open-boundary FFT solver on our
uniform mesh did not support the embedded conductor needed for the
holes. In WarpX we therefore used a multigrid solver for the resolved
array and the projected-source planar emitter, with homogeneous Neumann
boundary conditions on the electrostatic potential at the transverse faces
($\partial\phi/\partial n=0$, Appendix~\ref{app:numerics}).

At \SI{800}{nm} the rms energy spread for the projected-source planar emitter is
\SI{9.4}{\percent} larger in IMPACT-T than in WarpX, while the
projected normalized emittances are \SI{2.5}{\percent} smaller in $x$
and \SI{6.0}{\percent} smaller in $y$ (Appendix~\ref{app:flat_carriers}).
The codes also differ in mesh spacing and particle advance. We did not
conduct a cross-code convergence study and cannot separate the effect
of the boundary conditions from these other numerical differences.

With all 8192 samples per cell in the $41\times41$ array, we track about
$1.4\times10^7$ macroparticles from the projected-source planar emitter.
Before adopting sparser sampling for the injector, we
repeat this IMPACT-T calculation with 512 samples per
cell, about $8.6\times10^5$ macroparticles in total. We adjust their
weights to preserve the cell charge $q_c$ (Appendix~\ref{app:reduction}). The periodic WarpX
calculations still include all 8192 samples. Each retained particle
keeps its index $a$, so we can evaluate the correction for that same
sample at $q_c$ as in Sec.~\ref{sec:cell_charge}.

With this 16-fold reduction in particle count and the mesh coarsened
from $96\times96\times256$ to $48\times48\times128$, the whole-bunch
moments of the composed bunch at \SI{800}{nm} differ by at most
\SI{1.9}{\percent} from those with the higher IMPACT-T resolution.
The largest slice-profile
difference is \SI{2.3}{\percent} in rms energy spread.

\section{Application to an rf photoinjector}
\label{sec:injector_application}

We apply the composition method to a \SI{100}{pC} source in the
superconducting 1.6-cell, \SI{1.3}{GHz} rf gun studied for the
European XFEL~\cite{Bazyl2021}. In IMPACT-T we
include the gun and solenoid fields from the cathode onward. The solenoid
field peaks at \SI{0.19}{T}, \SI{0.51}{m} from the cathode. We retain the
nanohole geometry and local distributions of birth position and momentum
of Sec.~\ref{sec:model}. The \SI{0.87}{mm} diameter laser spot covers
about $1.1\times10^6$ holes. The Gaussian charge envelope, truncated
at $R=\sigma_r\simeq\SI{0.44}{mm}$, has an rms size of \SI{0.21}{mm}
per axis. On a flat cathode, initial kinetic energies $K_0$ uniformly
distributed from 0 to \SI{1}{eV} and isotropic outward directions give
a mean transverse energy
$\mathrm{MTE}=2\avg{K_0}/3\simeq\SI{0.33}{eV}$.
The normalized thermal emittance per unit rms spot size,
$\epsilon_{n,x}^{\mathrm{int}}/\sigma_x=\sqrt{\mathrm{MTE}/(m_ec^2)}$,
is \SI{0.81}{mm.mrad} per millimetre, giving \SI{0.17}{mm.mrad}
for this spot.

The emitted current has a flat-top temporal profile with a FWHM of about
\SI{13}{ps}. The sampled electron birth times span \SI{19.45}{ps}.
For a peak on-axis rf field of \SI{55}{MV.m^{-1}} and a phase
of about $40^\circ$ at the laboratory time origin, the applied field
magnitude at the cathode rises from \SI{26.0}{MV.m^{-1}} to
\SI{31.5}{MV.m^{-1}} during emission (Appendix~\ref{app:injector_field}).
Early electrons leave the periodic domain while emission continues.
For the planar periodic emitter and the structured periodic emitter,
we include the retarding field of these electrons and their image at the cathode
(Sec.~\ref{sec:long_emission}). This field peaks at
\SI{15}{MV.m^{-1}} at the spot center, against an applied rf field
of \SI{31}{MV.m^{-1}}, and at \SI{5.6}{MV.m^{-1}} at the outermost sampled radius
[Fig.~\ref{fig:departed_field}(b)].

In WarpX we track 8192 macroparticles per period on a
$64\times64\times160$ mesh with a \SI{1}{fs} step. The eight cell charges
span \SIrange{0.072}{0.12}{fC}. For the $41\times41$ array we reduced
the IMPACT-T particle count from 8192 to 512 per hole. At 512 per hole
the injector spot would require $5.5\times10^8$ macroparticles.
To limit tracking time, we retain one macroparticle
per hole for the projected-source planar emitter. Each carries the charge
emitted by its hole, about 450--740 electrons. The index identifying its
periodic trajectories varies from hole to hole so that all 8192 emission
samples are represented across the spot. With one sample per hole, the
charge-weighted rms corrections to $u_x$ and $u_y$ for the transmitted
samples across the spot at $H$ differ by at most \SI{0.01}{\percent} from those calculated
with all 8192 periodic samples at each hole's charge
(Appendix~\ref{app:injector_sampling}). We track electrons from the
projected-source planar emitter in IMPACT-T
with \SI{5}{fs} steps near the cathode and a
$96\times96\times256$ mesh for the bunch field and planar cathode image.

\begin{figure}[!htbp]
 \centering
 \includegraphics[width=0.96\linewidth]{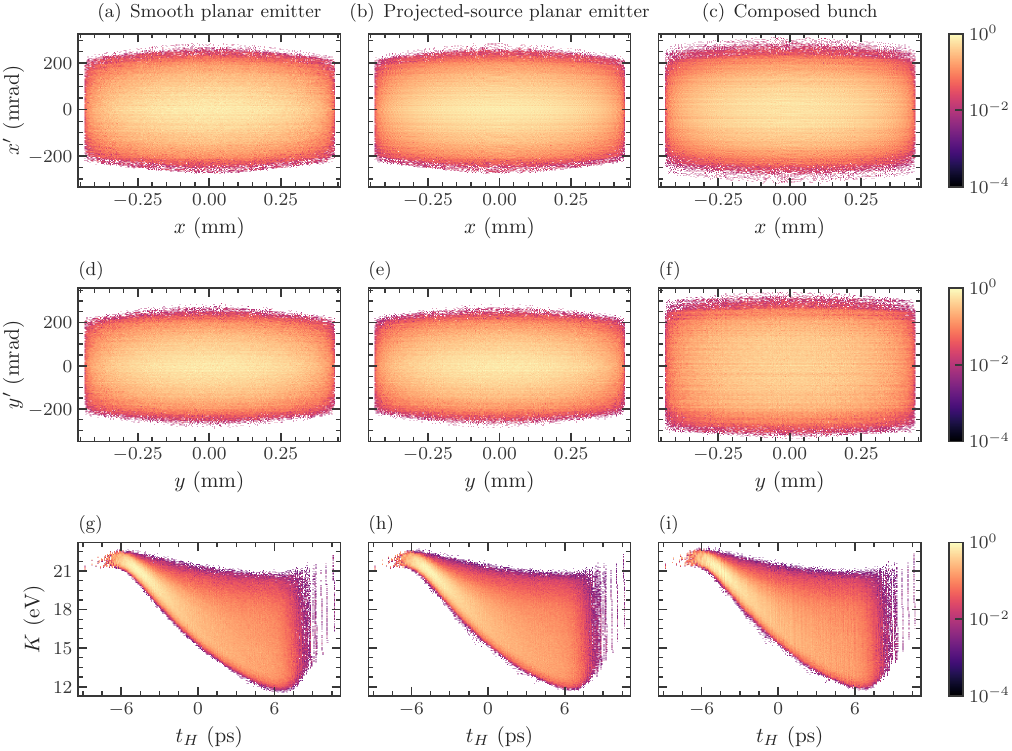}
 \caption{Injector phase space at \SI{800}{nm} for all transmitted
 particles. Panels (g--i) show the kinetic energy $K$ at crossing time
 $t_H$ on the laboratory rf clock. Color is the
 charge per bin relative to the largest bin charge in the same row,
 with one scale for all three bunches.}
 \label{fig:injector_handoff}
\end{figure}

For the \SI{0.87}{mm} diameter injector spot, we also track electrons
from an unpatterned flat cathode, the \emph{smooth planar emitter}.
Figure~\ref{fig:injector_handoff} shows all three bunches at $H=\SI{800}{nm}$.

The transmitted charge is \SI{100}{pC} for both the
smooth planar emitter and the projected-source planar emitter. For each, the rms
divergence is \SI{99}{mrad} in both transverse planes. For the composed
bunch, which represents emission from the nanostructured cathode, it
is \SI{110}{mrad} in $x$ and \SI{130}{mrad} in $y$.
Particles in the composed bunch reach $H$ on average \SI{43}{fs} later
than their counterparts from the projected-source planar emitter. The rms
arrival-time spread is \SI{3.8}{ps} for both bunches. From the trajectories
calculated for the structured periodic emitter we assign \SI{0.49}{pC}
to electrons that return to the cathode, leaving \SI{99.51}{pC} in the
composed bunch at $H$.

\begin{figure}[!htbp]
 \centering
 \includegraphics[width=0.96\linewidth]{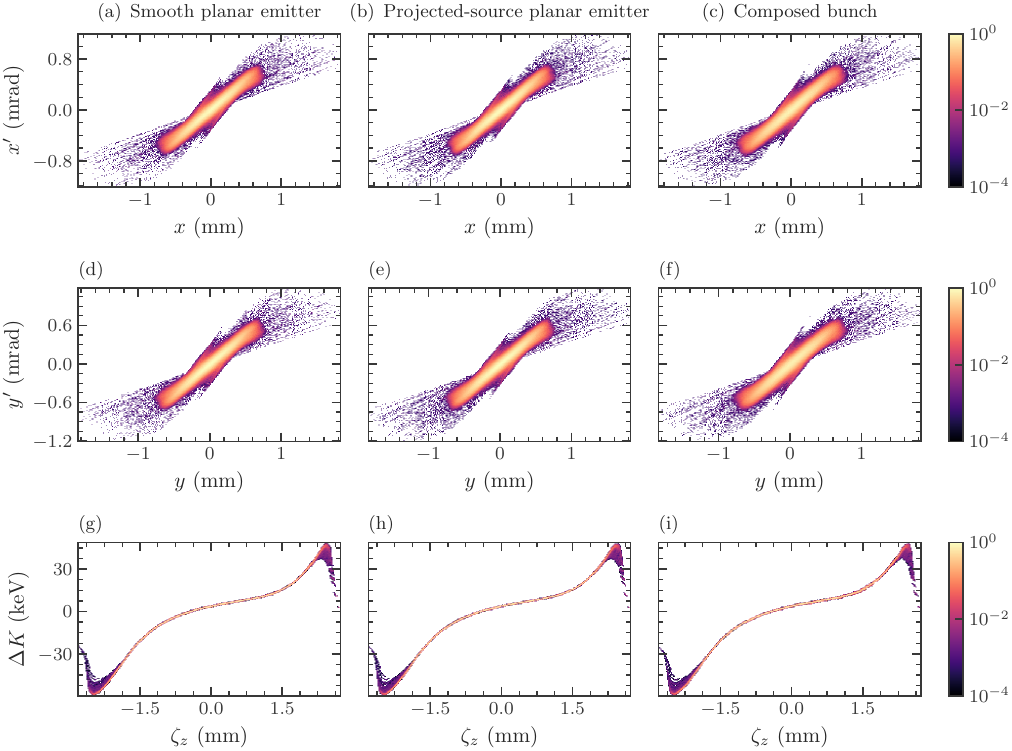}
 \caption{Whole-bunch phase space at \SI{5}{m}. Each row has a common
 color scale, as in Fig.~\ref{fig:injector_handoff}. The longitudinal
 position $\zeta_z$ and energy deviation $\Delta K$ are relative to the
 means for the bunch from the projected-source
 planar emitter.}
 \label{fig:injector_downstream_phase_space}
\end{figure}

We follow electrons from the smooth planar emitter and the projected-source
planar emitter continuously from emission. For the composed bunch we
start each trajectory at its corrected crossing time at $H$.
Electrons already emitted but still below $H$
therefore do not contribute to the downstream field. At the observation
planes \SI{1.52}{m} and \SI{5}{m} from the cathode
we record particle positions and momenta at a common time near the
passage of each bunch center. Figure~\ref{fig:injector_downstream_phase_space}
shows the phase space at \SI{5}{m}.

At each plane the central slice contains the middle 5\% of the transmitted
charge, ordered by longitudinal position. To check whether restarting
affects the central-slice emittances, we repeated the tracking for the
projected-source planar emitter using the uncorrected crossing states
at $H$. In either transverse plane the emittances differed from those
for continuous tracking of the same particles by less than 0.5\% at
\SI{1.52}{m} and by less than 0.6\% at \SI{5}{m}
(Appendix~\ref{app:timed_handoff}).

At \SI{5}{m} all three bunches have an rms length of \SI{1.2}{mm}
and similar transverse phase-space projections. No periodic modulation
is visible with transverse position bins \SIrange{14}{15}{\micro m} wide.
The longitudinal phase-space distributions are visually indistinguishable
at the resolution of Fig.~\ref{fig:injector_downstream_phase_space}(g--i).

\begin{figure}[!htbp]
 \centering
 \includegraphics[width=\linewidth]{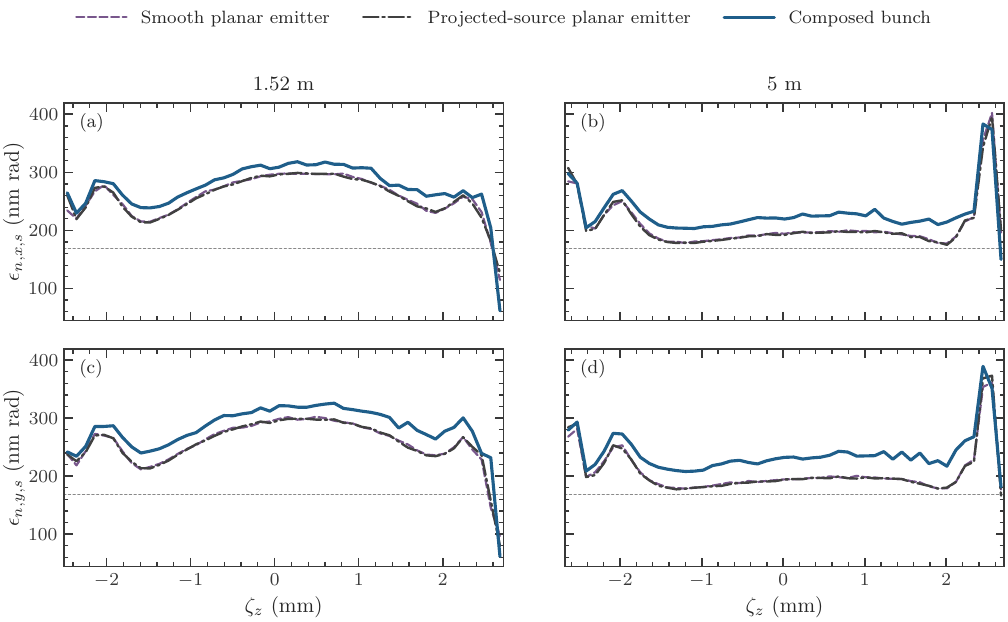}
 \caption{Normalized slice emittance in common longitudinal bins relative
 to the mean position of the bunch from the projected-source planar
 emitter. The dotted line marks the nominal planar thermal emittance,
 $\SI{170}{nm.rad}=\SI{0.17}{mm.mrad}$.}
 \label{fig:injector_slices}
\end{figure}

Over much of the bunch length, the slice emittances of the composed bunch
exceed those for the smooth planar emitter and the projected-source planar emitter
(Fig.~\ref{fig:injector_slices}). At \SI{1.52}{m} the central-slice
emittances of the composed bunch exceed those for the smooth planar
emitter by 4.0\% in $x$ and 6.5\% in $y$.

At \SI{5}{m} the normalized central-slice emittances of the composed bunch are
\SI{0.22}{mm.mrad} in $x$ and \SI{0.23}{mm.mrad} in $y$.
The vertical emittance exceeds the horizontal by 5.2\%. At the
cathode, emission is concentrated in two arcs on opposite sides of
each rim along $y$ [Fig.~\ref{fig:method}(a)], the direction of
greater divergence at $H$. For the smooth planar emitter the central-slice
emittances at \SI{5}{m} are about 15\% above the nominal planar thermal
emittance. Retaining the two arcs on a flat surface, as in the
projected-source planar emitter, changes either emittance by less than
2\%. For emission from the same arcs on the nanoholes, the predicted
central-slice emittances exceed those for the smooth planar emitter by 13\% in $x$
and 19\% in $y$ (Table~\ref{tab:injector_centers}).

\begin{table}[!htbp]
 \caption{Relative differences in normalized central-slice emittance.}
 \label{tab:injector_centers}
 \centering
 \small
 \renewcommand{\arraystretch}{1.14}
 \setlength{\tabcolsep}{3pt}
 \begin{tabular*}{\linewidth}{@{\extracolsep{\fill}}llcccc@{}}
  \toprule
  & & \multicolumn{2}{c}{\SI{1.52}{m}} & \multicolumn{2}{c}{\SI{5}{m}} \\
  \cmidrule(lr){3-4}\cmidrule(l){5-6}
  Bunch & Relative to & $x$ (\%) & $y$ (\%) & $x$ (\%) & $y$ (\%) \\
  \midrule
  Projected-source planar emitter & Smooth planar emitter & $-0.67$ & $-0.93$ & $-1.5$ & $-0.66$ \\
  \addlinespace[2pt]
  Composed bunch & Projected-source planar emitter & $+4.7$ & $+7.5$ & $+15$ & $+20$ \\
  \addlinespace[2pt]
  Composed bunch & Smooth planar emitter & $+4.0$ & $+6.5$ & $+13$ & $+19$ \\
  \bottomrule
 \end{tabular*}
\end{table}

At a second \SI{100}{pC} operating point with a smaller spot and a
longer pulse, we doubled the mean count from about one to two
macroparticles per hole without changing the mesh. At \SI{1.52}{m} the
projected emittances of the composed bunch decreased by less than
\SI{0.5}{\percent} in either plane. In a fixed \SI{0.15}{mm} slice
at the bunch center, the emittances decreased by \SI{0.63}{\percent} in $x$
and \SI{1.3}{\percent} in $y$. At both counts the central-slice
emittances remained higher than for the projected-source planar emitter
(Appendix~\ref{app:injector_sampling}).

\section{Discussion}
\label{sec:discussion}

In the $41\times41$ validation array, neighboring holes emit appreciably
different charges because the Gaussian charge envelope has an rms width
of only $6.5p$. Most of that charge originates on the sloping sides of
holes deeper than their nominal width. We calculate the correction at
each hole's own charge but assume that its periodic neighbors emit the
same charge. Even with this approximation, the projected emittances and
slice-emittance profiles of the composed bunch agree with the fully
resolved WarpX values to within \SI{0.1}{\percent} in both transverse
planes at both observation heights. The largest profile difference is in
the slice energy spread (Table~\ref{tab:planes}).

In the injector the charge envelope varies over about $580p$, so
neighboring holes emit nearly equal charges except at the outer edge.
The longer emission pulse introduces a different limitation. With an
emitted-current FWHM of \SI{13}{ps}, electrons leave the periodic
domain while the cathode is still emitting. We estimate their
field from trajectories calculated for the planar periodic emitter,
neglecting any change due to the hole geometry. The \SI{87}{fs} validation pulse does not test this
approximation. For the injector we have no fully resolved reference,
and we did not vary the particle count or mesh for the slice emittances
reported at \SI{5}{m}.

Above $H$ we no longer resolve the nanoholes. If the observation
plane is too close to the cathode, we omit part of the transverse
impulse from the distorted extraction field. For the \SI{747}{nm} pitch the
fundamental geometrical harmonic falls to \SI{3.5}{\percent} of its
surface amplitude by \SI{400}{nm} and to \SI{0.12}{\percent} by
\SI{800}{nm} in charge-free vacuum. A higher plane may be needed if
space charge within a beamlet still deflects electrons differently
in the two periodic emitters. Once those local deflections have
largely ceased, there is an advantage to keeping $H$ low. Below $H$
we calculate the bunch field with the projected-source planar emitter,
without the change in that field due to the nanoholes. With a lower
plane we neglect this change over a shorter distance and include each
particle's charge sooner in downstream tracking. For a new cathode
and emission regime, we would choose the lowest $H$ at which the
geometrical modulation of the extraction field is small and the transverse
momentum differences between matched periodic particles change little
when the plane is raised.

The agreement for the resolved array makes the method promising for
other periodically structured cathodes for which a photoemission
model supplies the charge, positions, times and momenta at emission.
In the present calculations these quantities are prescribed. If the surface field
changes the emission rate, for example through Schottky barrier
lowering~\cite{Bulgacheva2026}, electrons emitted earlier and their
cathode image can affect how much charge is emitted later in the
pulse~\cite{Omoumi2026}. The source and the time-dependent cathode
field would then have to be calculated together.

\section{Conclusion}

We formulated a composition method for tracking a bunch from a
periodically structured photocathode and validated it against direct
WarpX tracking for a fully resolved $41\times41$ array of nanoholes.
The composed bunch reproduces the angular broadening and the tail of
late arrivals. At the observation plane its projected normalized
emittances differ from those of the resolved bunch by at most
\SI{0.05}{\percent} in either transverse plane.

We applied the method to a \SI{100}{pC} source covering more than a
million holes and followed the composed bunch through the rf gun
and solenoid. At \SI{5}{m} no periodic modulation is visible in the
whole-bunch phase-space projections at the resolution shown. The
predicted central-slice emittances are 13\% higher horizontally and
19\% higher vertically than for the smooth planar emitter. The
central-slice emittances for the projected-source planar emitter, which
retains the hole emission pattern on a flat cathode, differ from those
for the smooth planar emitter by less than 2\% in both planes.
At this operating point most of the predicted increase is thus
associated with electron motion near the curved walls rather than
with the emission pattern alone.

\section{Acknowledgments}
The authors thank Klaus Floettmann for his valuable advice and insightful discussions. We also thank Erion Gjonaj and Margarita Bulgacheva from TEMF, TU Darmstadt for helpful regular exchanges on different approaches to modeling optical properties and beam dynamics of structured photocathodes. We thank Hans Weise for his interest and support. This work was performed as part of R\&D for future accelerator operation modes at the European XFEL
and financed by the European XFEL GmbH.

\appendix

\section{Emission and macroparticle sampling}
\label{app:source}

\subsection{Emission within one period}

In the cell $[-p/2,p/2)^2$ centered on a hole, set
$r=(\xi^2+\eta^2)^{1/2}$ and $\bar r=\min(r,p/2)$. For a Gaussian
$G(r)=\exp[-4\ln2\,(r/w_g)^2]$ of nominal FWHM $w_g$, the tracking
surface of depth $h$ is
\begin{equation}
 z_s=-h\,\frac{G(\bar r)-G(p/2)}{1-G(p/2)} .
 \label{eq:surface}
\end{equation}
The outward normal $\bm n$ defines the polar axis and the unit tangent
$\bm t_1$ along $(1,0,\partial_\xi z_s)$ defines zero azimuth. With
$\bm t_2=\bm n\times\bm t_1$, the projected initial proper velocity is
\begin{equation}
 \bm u_{0,\mathrm{flat}}=(\bm u_0\!\cdot\!\bm t_1)\bm e_x+
 (\bm u_0\!\cdot\!\bm t_2)\bm e_y+(\bm u_0\!\cdot\!\bm n)\bm e_z .
 \label{eq:flat_mapping}
\end{equation}

For the radial and angular intensity profiles in Fig.~4(b,c) of
Ref.~\cite{Li2013}, we use shape-preserving cubic interpolants
$\widetilde B$ and $C$. We rescale the radial coordinate to $[0,2L]$,
where $\ell(r)=\int_0^r\sqrt{1+h^2[G'(s)]^2}\,\dd s$ and $L=\ell(p/2)$,
and repeat $C$ with period $\pi$ without reflection:
\begin{align}
 B(r)&=\tfrac12\{\widetilde B[L-\ell(\bar r)]+\widetilde B[L+\ell(\bar r)]\},\nonumber\\
 I(r,\psi)&=B(r)\frac{C(\psi)}{C(\pi/2)}
       +\Lambda\chi^2(3-2\chi)\cos^2\psi ,
 \label{eq:intensity_surrogate}
\end{align}
where $\psi=\operatorname{atan2}(\eta,\xi)$,
$\chi=\max[0,(\bar r-r_{\rm ref})/(p/2-r_{\rm ref})]$ and
$r_{\rm ref}\simeq\SI{210}{nm}$. The second term extends the intensity
onto the surrounding plane.

The projected birth density is proportional to
$I^3\sqrt{1+|\nabla_\perp z_{\rm opt}|^2}$, where $z_{\rm opt}$ is the sum of
the central Gaussian and its eight neighbors, normalized to central depth $h$.
We choose $\Lambda\simeq1.88$ so that the area-weighted mean intensity
on a $768^2$ midpoint grid is one. We sample a $1024^2$ midpoint grid
using $z_s$ for birth heights and normals. Using $z_s$ also in the area factor changes
the normalized density by $5.4\times10^{-6}$ in total variation.

\subsection{Sampling the validation array}
\label{app:reduction}

We use the first $N_{\rm loc}=8192$ scrambled Sobol points (seed
2026082801). One coordinate selects a grid element by cumulative birth
density, two give a uniform position within it, and four set $K_0$,
$\cos\theta$, $\varphi$ and $t_b$. The Gaussian birth
times have rms width $\sigma_t=T_L/(2\sqrt{6\ln2})\simeq\SI{37}{fs}$
for laser-intensity FWHM $T_L$, with truncation at $t_c\pm4\sigma_t$
and $t_c=4\sigma_t+\SI{1}{fs}$. For the resolved array and projected-source planar emitter,
$w_{ca}=p^2\Sigma_q(\bm R_c+(\xi_a,\eta_a))/(eN_{\rm loc})$, where
$\Sigma_q(\bm r)=\Sigma_0\exp[-|\bm r|^2/(2\sigma_q^2)]$.
In both periodic emitters the weight is $q_j/(eN_{\rm loc})$, with
$q_j/(p^2\Sigma_0)$ in
$\{1,0.8,0.6,0.4,0.2,0.05,0.01,5\times10^{-5}\}$.

For the projected-source planar emitter with 512 samples per cell in
IMPACT-T, let $W_c=\sum_aw_{ca}$. For each
$k=0,\ldots,511$ retain the first index $a$ satisfying
$\sum_{b\le a}w_{cb}>(k+\vartheta_c)W_c/512$, where
\begin{equation}
 \vartheta_c=\operatorname{frac}\big[(i_x+20.5)(\sqrt2-1)
 +(i_y+20.5)(\sqrt3-1)+0.125\big].
 \label{eq:cell_phase}
\end{equation}
Here $(i_x,i_y)$ index cells relative to the center and
$\operatorname{frac}$ denotes the fractional part. Retained weights are
$W_c/512$, with the last adjusted for roundoff.

\subsection{Sampling the injector sources}
\label{app:injector_sampling}

We index cell centers within $R=\sigma_r\simeq\SI{436}{\micro m}$ from
$c=0$ by increasing $i_x$, then $i_y$. For the projected-source planar
emitter the sample index is $(2473c+1)\bmod8192$ (zero-based) and its
weight is $q_c/e$, with $q_c\propto\exp[-|\bm R_c|^2/(2\sigma_r^2)]$
and $\sum_c q_c=\SI{100}{pC}$. For the smooth planar emitter we replace
only the transverse birth positions with independently permuted samples
of the circular Gaussian truncated at $R$. For both emitters we map
Sobol time ranks to quantiles of $10^4$ tabulated birth times (pulse
length parameter $L_t=\SI{12.65}{ps}$, edge parameter
$r_t=\SI{2}{ps}$). Births span $-9.45$ to $+10.00$~ps, with an
emitted-current FWHM of about \SI{12.65}{ps}.

For $\sqrt{\langle(\Delta u_d)^2\rangle_q}$ at $H$ ($d=x,y$), the
charge weight per transmitted sample is $q_c$ for one sample per cell
and $q_c/8192$ for the full set.

The particle-count check used $2^{19}$ and $2^{20}$ macroparticles over
the same $5.2\times10^5$ holes (\SI{0.30}{mm} radius, birth-time
span about \SI{28}{ps}). At \SI{1.52}{m} the common
\SI{0.15}{mm} central slice contains the mean $z$ of the higher-count
bunch from the projected-source planar emitter. The central-slice
emittances of the composed bunch exceeded those for the projected-source
planar emitter by $(3.1,4.5)\%$ in $(x,y)$ at the lower count and
$(2.8,5.1)\%$ at the higher count.

\section{Particle tracking and beam diagnostics}
\label{app:numerics}

\subsection{Fields and trajectories in the validation calculations}

The transverse WarpX domains are $45p\times45p$ for the resolved array
and the projected-source planar emitter (flat margin $2p$ on each side)
and $p\times p$ for either periodic emitter. The lower boundary is
$z=-p/2$ in both cases. The upper boundaries are $5p/2$ and $2p$, respectively.
The cathode and lower face are grounded. The upper potential
is $E_0z_{\rm hi}$ for extraction field $E_0$. The transverse potential
conditions are homogeneous Neumann on the $45p\times45p$ domain and
periodic on the $p\times p$ domain. Nonperiodic faces absorb particles.
Optical-frequency forces, magnetic self-fields and retardation are omitted.
Table~\ref{tab:cost} gives the mesh sizes, particle counts and run times.

\begin{table}[!htbp]
 \caption{Execution costs for the validation array. The IMPACT-T entries
 are for the projected-source planar emitter. For the planar periodic
 emitter and the structured periodic emitter, counts are per run and
 times and costs are totals over all 16 runs.}
 \label{tab:cost}
 \centering
 \small
 \renewcommand{\arraystretch}{1.12}
 \setlength{\tabcolsep}{3pt}
 \begin{tabular*}{\linewidth}{@{\extracolsep{\fill}}lccccc@{}}
  \toprule
  Calculation & Macroparticles & Mesh & Nodes &
  \shortstack{Wall time\\(min)} & \shortstack{Cost\\(node h)} \\
  \midrule
  IMPACT-T & $1.4\times10^7$ & $96^2\times256$ & 24 & 13 & 5.3 \\
  IMPACT-T & $8.6\times10^5$ & $96^2\times256$ & 24 & 3.3 & 1.3 \\
  IMPACT-T & $8.6\times10^5$ & $48^2\times128$ & 8 & 1.3 & 0.17 \\
  \addlinespace[3pt]
  WarpX, resolved array & $1.4\times10^7$ & $2880^2\times192$ & 24 & 220 & 87 \\
  \parbox[c]{11em}{\raggedright WarpX, projected-source\\planar emitter} & $1.4\times10^7$ & $2880^2\times192$ & 24 & 180 & 71 \\
  WarpX, periodic emitters & 8192 & $64^2\times160$ & 1 & 74 & 1.2 \\
  \bottomrule
 \end{tabular*}
\end{table}

In WarpX we use first-order charge deposition, one binomial filter pass,
momentum-conserving field interpolation and a relativistic Boris advance.
The multigrid relative residual tolerance is $10^{-7}$ (1000 iterations
maximum). At birth we extrapolate the vacuum field linearly to the surface
from $1.25\Delta x$ and $2.25\Delta x$ along the outward normal and advance
the electron for the remainder of the step. Newly emitted electrons
moving outward within $0.5\Delta x$ of the surface are excluded from
absorption. Returns require
flight time $>\Delta t/2$ and displacement from birth $>\Delta x/4$.
In the injector periodic cells we also require inward normal momentum
and positive kinetic energy.

In WarpX we find the first upward crossing of $H$ by cubic Hermite
interpolation of the bracketing positions and velocities ($v_z>0$).
Time and proper velocity are interpolated linearly. In IMPACT-T V3.0
we compute the bunch and planar image fields by convolution on a doubled
FFT grid with cloud-in-cell deposition and interpolation. We use a
drift--kick--drift step with a relativistic Boris kick and interpolate
the first upward crossing linearly after the drift half-step.
\subsection{Particle matching and cathode returns}
\label{app:response_assignment}

At $q_c=q_j$ we use the tabulated difference without interpolation.
Nearest-charge ties go to the higher $q_j$. There is no extrapolation
or redistribution of returned charge. For $H=\SI{800}{nm}$, particles
identified as cathode returns both by direct tracking of the resolved
array and by the composition method account for \SI{0.24}{\percent} of
the emitted charge. In each case less than \SI{0.007}{\percent} had
neither crossed $H$ nor returned to the cathode by the final time.

\subsection{Timed injector restart}
\label{app:timed_handoff}

For $t_i=t_i^{\rm comp}(H)$, a particle is activated at
\begin{equation}
 t_i^{\rm act}=t_0+\Delta t
 \left\lceil\frac{t_i-t_0}{\Delta t}\right\rceil,\qquad
 t_0=\min_i t_i,\quad \Delta t=\SI{5}{fs}.
 \label{eq:injector_activation}
\end{equation}
Before activation, position and momentum remain fixed and no charge is
deposited. Beyond the near-cathode segment we use a \SI{0.1}{ps} time
step and omit the cathode image field. No particle in the composed
bunch recrossed $H$ or was lost.
For the restart check with the projected-source planar emitter, we used
the uncorrected crossing states at $H$ and restarted at \SI{1.52}{m}
with saved positions, momenta and weights
on the same rf clock.

\subsection{Beam diagnostics}

The normalized emittance is $\epsilon_{n,x}=\sqrt{\operatorname{var}_w(x)
\operatorname{var}_w(u_x)-\operatorname{cov}_w(x,u_x)^2}/c$ and likewise
in $y$, with physical particle weights for the transmitted bunch or
the selected slice. For bin charge $Q_k$, the current magnitude is
$Q_k/\Delta t_k$ at a crossing plane or
$Q_k\langle v_z\rangle_{w,k}/\Delta z_k$ at a common time.

In Fig.~\ref{fig:phase_space}, $180\times180$ transverse bins span
$\pm4.25$ rms widths of the resolved bunch and $220\times138$ longitudinal bins cover
the range of the matched particles. For bin charge fractions $f_k$, the total
variation distance $\tfrac12\sum_k|f_k-f_{k,\rm res}|$ in the three rows
is $(0.16,0.30,0.41)$ for the projected-source planar emitter and
$(0.01,0.01,0.06)$ for the composed bunch.

For Table~\ref{tab:planes} we retain the same 50 nearly equal-charge groups
of matched particles, sorted by arrival time in the resolved bunch.
For a profile $\bm A$ the relative difference is
\begin{equation}
 D_A=\frac{\|\bm A^{\rm cand}-\bm A^{\rm ref}\|_2}{\|\bm A^{\rm ref}\|_2}.
 \label{eq:slice_error}
\end{equation}
For IMPACT-T we define the equal-charge intervals from the composed
bunch with 8192 particles per cell and bin each transmitted particle by
its own arrival time. Figures~\ref{fig:slices}, \ref{fig:practical}
and~\ref{fig:flat_carriers} show profiles in common fixed-width time bins.
For energy and emittance only, we apply a three-point mean (available
bins at the ends) and shape-preserving interpolation. In Fig.~\ref{fig:practical} these
moments are omitted below 0.5\% of the peak current of the resolved bunch.

The injector phase-space histograms have common $256\times256$ bins.
For Fig.~\ref{fig:injector_slices} we use 50 common longitudinal bins per
plane and omit bins below 0.5\% of the projected-source planar emitter's
peak current. For the central-slice emittances in Table~\ref{tab:injector_centers},
we sort by $z$ and select particles with cumulative charge fractions
$(\sum_{k<i}w_k+w_i/2)/\sum_k w_k$ between 47.5\% and 52.5\%.

\subsection{Projected-source planar emitter in WarpX and IMPACT-T}
\label{app:flat_carriers}

\begin{figure}[!htbp]
 \centering
 \includegraphics[width=0.85\linewidth]{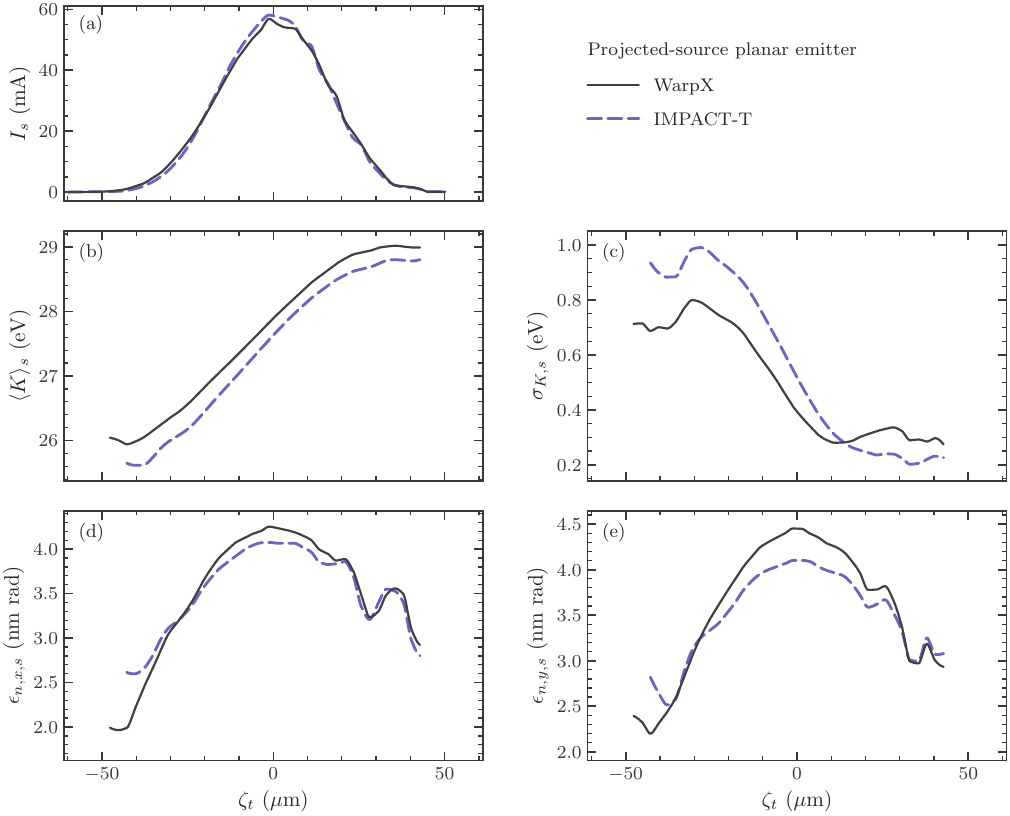}
 \caption{Slice profiles at \SI{800}{nm} for the projected-source planar
 emitter with 8192 macroparticles per cell. Arrival times are relative to the WarpX mean.
 Bins below 0.5\% of the WarpX peak current are omitted from the moment curves.}
 \label{fig:flat_carriers}
\end{figure}

At \SI{800}{nm} (Fig.~\ref{fig:flat_carriers}), the whole-bunch rms
energy and arrival-time spreads are \SI{0.87}{eV} and \SI{51}{fs} in
WarpX, and \SI{0.95}{eV} and \SI{49}{fs} in IMPACT-T. The respective
projected normalized emittances $(\epsilon_{n,x},\epsilon_{n,y})$
are $(3.99,4.14)$ and $(3.89,3.89)$~nm\,rad.

\FloatBarrier
\section{Field reconstruction over the finite emitting area}
\label{app:injector_field}

On the laboratory clock $E_{z,\rm RF}(0,t)=-\mathcal E_{\rm RF}(t)$
at the planar cathode, with $\mathcal E_{\rm RF}(t)=E_c\sin(2\pi f t+\phi_0)$,
$E_c\simeq\SI{45}{MV.m^{-1}}$, $f=\SI{1.3}{GHz}$ and
$\phi_0\simeq39.86^\circ$. The WarpX time is $t_{\rm W}=t-t_{b,\min}$.
In IMPACT-T the local time $t-t_0$ for a segment starting at $t_0$
requires cosine phase $90^\circ+\phi_0+360^\circ f t_0$.

For \SI{40}{ps} of periodic injector tracking, the space-charge potential
obeys $\phi_{\rm sc}=0$ at the cathode and $\partial_z\phi_{\rm sc}=0$
at $z_{\rm hi}=2p$. Both longitudinal faces absorb particles.
At iteration $n$ and charge $q_k$ the scale factor for the charge-free
\SI{35}{MV.m^{-1}} reference field is
$(\mathcal E_{\rm RF}-g_{n,k})/(\SI{35}{MV.m^{-1}})$.
For electrons from the planar periodic emitter, we synchronize the
half-step momentum to departure at $z_{\rm hi}$ and advance $z_{ka}(t)$
in the longitudinal rf electric field
alone (relativistic kick--drift--kick, steps $\le\SI{25}{fs}$).

For $q(r)=q_8\exp[-r^2/(2\sigma_r^2)]$, the linear weights $\alpha_k(r)$
of the two bracketing charges sum to one. Each of the eight equally
spaced $q_k$ represents $M_k=\sum_c\alpha_k(|\bm R_c|)$ cells.
For a ring of radius $r'$ at height $z$, the normal cathode field per
unit ring charge magnitude, including its planar image, is
\begin{equation}
 K_{\rm ring}(r,r',z)=
 \frac{z\,\mathrm E(m)}{\pi^2\epsilon_0
 \sqrt{(r+r')^2+z^2}\,[(r-r')^2+z^2]},
 \qquad m=\frac{4rr'}{(r+r')^2+z^2},
 \label{eq:ring_kernel}
\end{equation}
where $\mathrm E(m)$ is the complete elliptic integral of the second
kind with parameter $m$. For charge $q_k$ the radial average is
\begin{equation}
 K_k(r,z)=\frac{\int_0^R r'\alpha_k(r')K_{\rm ring}(r,r',z)\,\dd r'}
                  {\int_0^R r'\alpha_k(r')\,\dd r'} .
 \label{eq:radial_basis_kernel}
\end{equation}
At time $t$, $D_k^{(n)}(t)$ includes the electrons from the planar
periodic emitter that have left through $z_{\rm hi}$ and not returned
to $z=0$. At radius $r_j$
with $q(r_j)=q_j$, the next field is
\begin{equation}
 g_{n+1,j}(t)=\sum_{k=1}^{8}\frac{M_kq_k}{N_{\rm loc}}
 \sum_{a\in D_k^{(n)}(t)} K_k(r_j,z_{ka}(t)).
 \label{eq:returned_field_iteration}
\end{equation}
Linear interpolation of $g_2$ and $g_3$ to the sampled birth times gives
\begin{equation}
 \epsilon_g=\frac{\|g_3-g_2\|_{\rm em}}{\|g_3\|_{\rm em}}
 \simeq4.8\times10^{-4},\qquad
 \|g\|_{\rm em}^2=\sum_{j,a}M_jq_j\,g_j^2(t_{b,a}).
 \label{eq:returned_field_difference}
\end{equation}

\setlength{\bibsep}{-2.5pt}
\bibliography{references}

\end{document}